\documentclass[twocolumn,showpacs,amsmath,prl]{revtex4}

\usepackage{graphicx}
\usepackage{dcolumn}
\usepackage{bm}

\begin{document}

\title{Charge ordering in oxides : a conundrum solved by
resonant diffraction }

\author{Y. Joly}
\email{yves.joly@grenoble.cnrs.fr}
\author{S.~Grenier}
\altaffiliation[Present address: ]{Dept. of Physics, Brookhaven National
Laboratory, Upton, NY 11973, USA. \texttt{grenier@bnl.gov}}
\author{J.E.~Lorenzo}
\email{emilio.lorenzo@grenoble.cnrs.fr}
\affiliation{Laboratoire de Cristallographie, CNRS, B.P.~166, F-38042 Grenoble Cedex 9, France.\\}

\date{\today}

\begin{abstract}
We show that in mixed-valence $3d$ transition metal oxides
undergoing a structural transition, the low temperature phase
results from an effective ordering of the charge. This arrangement
and the quantitative evaluation of the atomic charges are
determined by using resonant x-ray scattering experiments further
analyzed with the help of {\it ab initio} calculations of the
corresponding scattering factors. We have found that this
reorganization concerns only a small fraction of electron and is
necessary to reconcile all the experimental data.

\end{abstract}

\maketitle


There is an issue in understanding the charge state in the high
and low temperature phase of mixed-valence transition metal
compounds. Few experimental techniques can access the actual
charge state of a given atom and most of the accumulated knowledge
results from theoretical models \cite{1} and from indirect results
based on transport property measurements \cite{2} or from
heuristic bond-valence calculations on the crystal structure
\cite{3}. Nevertheless, the answer to very simple questions would
turn out to be of large relevance in the context of
metal-insulator or insulator-insulator phase transitions issued
from a charge ordering (CO), chiefly if one could deduce a general
pattern. For instance, given an average valence state as deduced
from charge neutrality considerations of the chemical composition,
would the ground state be an effective CO ? and if so
how are the corresponding electrons arranged ? Are they located in
the very localized $3d$ orbital of the transition metal, or in the
extended $4s$ or even on the oxygen ligands ? In this paper we
claim that resonant X-ray diffraction is able to answer
quantitatively these questions. In our study, we have found a
small but effective ordering occurring mainly in the $3d$
transition metal orbitals. We think that the same trend occurs in
many compounds exhibiting CO which range in diversity
and physical properties from colossal magnetoresistance
manganites, to the astonishingly complex Verwey  \cite{4}
transition in magnetite, to the CO inducing
Spin-Peierls phase transition in NaV$_{2}$O$_{5}$  \cite{5}.

X-ray absorption spectroscopies (XAS) and resonant X-ray
scattering are extensively used techniques to explore and
elucidate the electronic properties of materials \cite{6,7}. In
these spectroscopies the incoming photons are absorbed by
promoting a core electron to an empty intermediate excited state
$f$. When the electron subsequently decays to the same core-hole,
it emits a second photon with the same energy than the incoming
one. This process is virtual and preserves the coherence of the
electromagnetic wave.  Set in diffraction conditions the
associated experimental technique is known as diffraction
anomalous fine structure spectroscopy or resonant diffraction.
When the absorption process is considered alone, and therefore
loosing the coherence properties of the X-ray field, the experimental
technique is called XAS or X ray absorption near edge spectroscopy
(XANES) when considering the structure close to the edge. In this
low energy range a higher sensitivity to the
electronic configuration around the photoabsorber is expected
because the photoelectron probes allowed electronic states very
close to the Fermi level as required by the selection rules
resulting from the Fermi golden rule. Indeed by playing with the
polarization of the incoming (and eventually outgoing) photon
beam, with the different channels of transition (dipole,
quadrupole) and with the orientation of the crystal with respect
to the photon polarization direction, the probability of
transition becomes highly angle dependent. This phenomena is
illustrated by a tensor form of the scattering structure factors,
and the analysis of each term of the tensors gives access to
specific projections of the electronic states. Although complete
calculations are performed we will restrict to the dipole approximation and the atomic scattering factor (ASF) reads,
\begin{eqnarray}
f'+if'' & = &  \sum_{\alpha\beta}\epsilon_{\alpha}^{o*}
\epsilon_{\beta}^i D_{\alpha\beta}
\end{eqnarray}
\noindent $\alpha$ and $\beta$ are the index of the coordinates in
the orthogonal basis. $\epsilon^{i}$ and $\epsilon^{o}$ stand for
the incoming and outgoing polarizations and $D_{\alpha\beta}$ is
given by:
\begin{eqnarray}
 D_{\alpha\beta}& =&  -m_e\omega^2 \sum_{fg} \frac{\langle \psi_g|r_{\alpha}|\psi_f\rangle
  \langle \psi_f|r_{\beta}|\psi_g\rangle}{E_f-E_g-\hbar\omega+i\frac{\Gamma(E_f)}{2}}
  \label{eq_D}
\end{eqnarray}
\noindent where $E_f$, $E_g$ and $\hbar\omega$ are respectively the
energies of the intermediate state, the core state and the photon; $m_e$ is
the electron mass and $\Gamma(E_f)$ includes the contribution of the inverse lifetime of all the states involved in the resonant process. The
summation over $f$ is limited to the unoccupied states. In
diffraction, the intensity of the reflections is proportional to
the modulus squared of a weighted sum of the ASF:
\begin{equation}
I(\vec{Q}; \hbar\omega) \propto \left|\sum_{a}e^{i\vec{Q}\cdot\vec{R_a}}\left(f_{0a}(\vec{Q})+f_a'(\omega)+if_a''(\omega)\right)\right|^{2}
\end{equation}

\noindent where $\vec{R_a}$ stands for the atom $a$ position and
$\vec{Q}$ is the diffraction vector. The atomic scattering
amplitude is the sum of the Thomson factor $f_{0a}$ and the ASF.

 Of particular
importance are the \emph{forbidden} Bragg peaks that are extinct
off resonance conditions because of the presence of glide planes
or screw axis, if allowed by the space group of the given
material. For these reflections, the diffracted intensity is
related to the difference between ASF of the resonant (or
anomalous) atoms. Therefore, these reflections are supposed to be
much more sensitive to slight modifications of the
crystallographic as well as the local electronic structure. In the
absence of new superlattice peaks originating from a phase
transition (typically ferrodistortive phase transitions), these
\emph{forbidden} Bragg peaks might yield relevant informations on
the corresponding CO.\\

This phenomenology has been extensively used in the study of the
interplay between orbital, spin and charge degrees of freedom in
transition metal oxides \cite{8,9}. In particular it is specially
suited to study charge \cite{10} and orbital orderings
\cite{11,12,13}. Nevertheless apparently contradictory results
have led to opposite conclusions on the presence or not of real
CO. Recent resonant diffraction experiments at \emph{forbidden}
reflections in Fe$_3$O$_4$ have yielded no change in the resonant
signal below and above the ordering temperature and it has been
concluded that no CO occurs at the Verwey transition \cite{14}.
Conversely, resonant scattering experiments in NaV$_{2}$O$_{5}$
\cite{15} and La$_{1/2}$Sr$_{1/2}$MnO$_3$ \cite{16} have shown
that the energy line shape of some reflections retains close
similarities with the derivative of the resonant X-ray scattering
factor, a fact which has been interpreted as a signature for the
presence of two different V (or Mn) charge states ordered in the
lattice. In this letter we shall show that all these experimental
evidences can actually be interpreted within a unique framework.
In order to accomplish this task, we shall proceed in a
quantitative analysis of the resonant diffraction experiments with
the help of \emph{ab initio} simulations. It will then be possible
to evaluate the influence of the different parameters and estimate
the effective charge
redistribution resulting from the CO transition. \\

To illustrate our analysis we shall discuss very recent resonant
X-ray diffraction results in NaV$_{2}$O$_{5}$ and all the
experimental details can be found in ref.~\cite{15}. Among the
numerous diffraction peaks measured as a function of the incident
photon energy and under different photon polarization conditions,
we select three of them to illustrate our results. The first one
is the (1,~0,~0)$_c$ \emph{forbidden} reflection which does not
exhibit any change through the transition.
The incoming photon
polarization is along $c$, the outgoing one is in the scattering
plane ($\sigma-\pi$ polarization condition). The two other
reflections under consideration are superlattice peaks,
(15/2,~1/2,~1/4)$_{b,c}$, for which the incoming and outgoing
photon polarizations were kept parallel ($\sigma - \sigma$
condition). According to the index, the polarization directions
are along either the $b-$ or $c-$axis. The superstructure peaks
appear due to the phase transition and the energy dependence of
the scattered intensity exhibits peculiar features that we examine
in detail below. \emph{Ab initio} monoelectronic calculations are
performed using the FDMNES code. \cite{17,18} Starting from the
atomic positions and an associated electronic density, this code
solves the Schr\"{o}dinger or Dyson equations to determine the
intermediate states probed by the photoelectron. The resulting
structure factors are evaluated by taking into account the
different excitation-deexcitation channels (dipole-dipole,
dipole-quadrupole and quadrupole-quadrupole). Diffracted
intensities are calculated with the crystallographic structures
corresponding to the high ($Pmmn$ \cite{19}) and low ($A112$
\cite{20}) temperature phases with and without CO. Because the
purpose of the present paper is to fit the electronic structure,
it is important to define easy to handle parameters. We choose the
valence orbital occupancy rates (VOOR). The radial shape of the
atomic orbitals remains fixed. For any VOOR set, the total
electronic density is obtained by simple superposition of the
electronic orbital densities weighted by the VOOR. First-guess
VOOR values corresponding to the high temperature (HT) phase are
obtained by a fit on the electron density calculated by a LAPW
\cite{21} band structure procedure. These orbital occupancies,
which reflect the location of the charge in space, are further
refined in order to accurately reproduce the experimental curves
in the low temperature phase. The vanadium and oxygen VOOR are
then the only adjustable parameters of our calculations. The HT
starting values are respectively 2.3 and 1.7 electrons for the
$3d$ and $4s$ V atoms and 4.6 electrons for the $2p$ O atoms. If
one uses the ionic radii as a definition of the atomic spheres,
the $4s$ charge (average radius is 1.5\AA, to be compared with the
$3d$ average radius, 0.6\AA) lies inside the oxygen atoms. Thus,
the V formal charge is 2.7+ in this context. The sodium atom has
no electron on its $3s$ valence orbital and remains 1+ in all the
study. With this VOOR set, a good agreement is obtained at HT
between the calculated (100) reflection, the polarized XANES line
shape  and the corresponding experimental spectra. To get such
agreement, the finite difference procedure is used to solve the
Schr\"{o}dinger equation. This technique permits to avoid the
muffin-tin approximation on the potential shape. That point is
absolutely necessary in such distorted systems.\\

\begin{figure}
\includegraphics[scale=0.38]{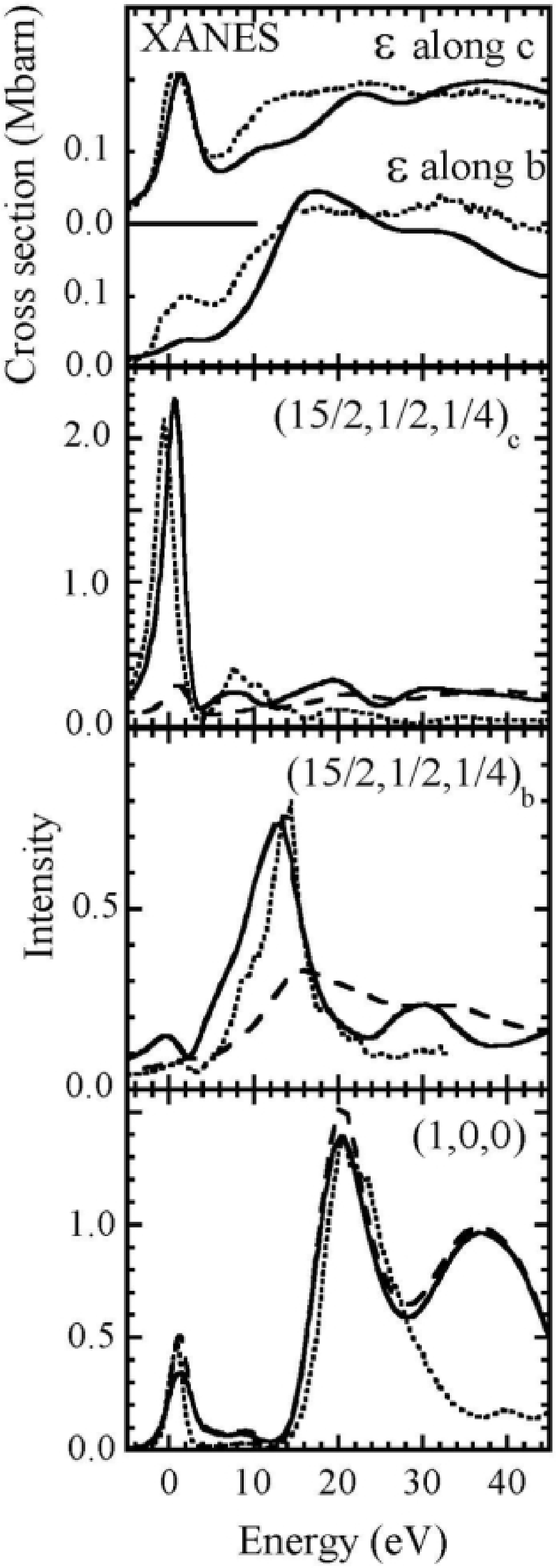}
\caption { \label{fig:fig1} Experimental and calculated resonant
diffracted peak spectra in NaV$_2$O$_5$. From bottom to top the
(1,~0,~0) $\sigma - \pi$, (15/2,~1/2,~1/4)$_{c}$ and
(15/2,~1/2,~1/4)$_{b}$ peaks (both measured in the $\sigma -
\sigma$ configuration) and the XANES cross section are shown. For
each reflection the spectra corresponding to the experiment
(dotted line) and the calculation in the low temperature phase
with charge ordering (continuous line) and without charge ordering
(dash line) are displayed. For the (1,~0,~0) reflection the high
temperature structure calculations are not shown as they are
practically identical to the LT calculation without charge
ordering. The charge order model yields very few modifications of
the signal through the transition for the first peak. The
derivative effect for the two last ones can not be accounted for
by a lattice distortion alone, and is best reproduced when
different charges are introduced at convenient V-sites. The XANES
spectra for the polarization $\epsilon$ along the $c-$ and
$b-$axis are also given. The different calculations at HT and LT
with or without charge ordering yields practically the same
results.}
\end{figure}

\begin{figure}
\includegraphics[scale=0.18]{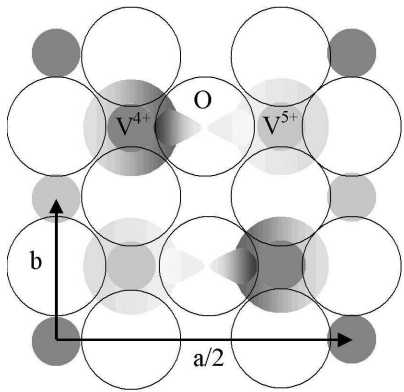}
\caption {\label{fig:fig2} A sketch of the NaV$_2$O$_5$ structure
focussed around the V$^{4+}$-O-V$^{5+}$ rungs. All oxygen atoms
lie in the same plane, whereas vanadium rungs lie alternatively
below and above the oxygen plane. The vanadium (full circles) are
surrounded by five oxygen atoms, four shown in white circles on
the figure and a fifth one in the apical direction (not shown).
The rings show the region were the electronic density is affected
by the CO, the difference in density around the vanadium atoms
being illustrated by the different grey tones. }
\end{figure}

We now discuss the low temperature (LT) phase and the starting
point is the crystallographic structure as determined in ref.~\cite{19}.
The most
important result is that an explicit CO is absolutely
necessary to get a satisfactory agreement with the experiment
(Fig.~\ref{fig:fig1}). The fit of the experimental data yields two
equivalent solutions : \emph{(a)} The charge exchange is located at the
$4s$ vanadium and amounts $\delta =\pm$ 0.2 electron, and \emph{(b)}
it is located at the $3d$ vanadium and
amounts only $\delta = \pm$ 0.04 electron. Thus the CO takes place either
in the intermediate region between the
oxygen and the vanadium where an atomic assignment of the electron
is merely formal (see Fig.~\ref{fig:fig2}) or closer to the atom
centers with a smaller magnitude. In any case, the bridging oxygen
located in the rung of the ladder, being the atom that experiences the largest
displacement at the phase transition (0.05\AA~towards the V$^{5+}$,
that is 5 to 10 times more than the other atoms), is suspected to participate
to the electronic rearrangement as well. Thus, the electronic
reorganization we propose does not concern isolated atoms but all
the rung of the ladder, with the center of gravity of the electronic
density displaced in the opposite direction than that of the
bridging oxygen, i.e., toward the V$^{4+}$ atom. Taking into
account the experimental resolution, the choice among the two
proposed solutions cannot be safely done. Nevertheless the second
one with a CO on the $3d$ orbital is in agreement
with that obtained by Suaud and Lepetit \cite{22} on the basis of
{\it ab initio} calculations using large configurations
interaction method on embedded fragments. These authors found a
$\pm$ 0.04 CO on the $3d$, in excellent agreement
with the present study. They have also shown that the modification of
the magnetic properties at the phase transition is due to the
redistribution between the different $d-$orbitals, keeping the
total $d$ charge only slightly modified. We want to end this
paragraph by underlying the extreme sensitivity of resonant
diffraction to observe such a small amount of charge
disproportionation.\\

\noindent

A second important result, with straightforward implications in
other CO compounds, is that the CO has very little influence on
the (1,~0,~0) \emph{forbidden} reflection whereas it is at the
origin of the derivative line shape for the non integer
reflections. A charge disporportionation in the vicinity of the
absorbing atom greatly affects localized orbitals, and thus both
the $1s$ and the $3d$ energy levels will undergo an analogous
shift. However the $4p$ states, less localized, are hardly
affected \cite{joly}. Therefore the dipole-dipole contribution to
the $K-$edge of an atom experiencing a CO will be shifted towards
higher or lower energies as compared to that found in the HT phase
whereas the quadrupole pre-edges are little shifted during the
oxidation processes. In a more general way, one has to proceed
very cautiously on applying the derivative effect concept to $L-$
and $M-$edges if the CO is expected to take place in $d-$ and $f-$
final states, respectively. We have computed the effect of the
associated geometrical distortion onto the ASF and we have
concluded that it remains negligible ($\pm$ 0.1 eV) as compared to
the observed energy shift and to the effect of the charge
disproportionation ($\delta E = \pm 0.8$ eV). A first order Taylor
expansion of the scattering factor is justified by the small value
of $\delta E$. It reads as $f^{+\delta} \approx f +
\partial f/\partial E\cdot\delta E$ and $f^{-\delta} \approx f -
\partial f/\partial E\cdot\delta E$ where the $f$, $f^{+\delta}$ and $f^{-\delta}$ are the
ASF tensors for the vanadium atoms with the HT charge, and the
$+\delta$ and $-\delta$ charge disproportionation, respectively.
The derivative effect condition is best achieved for diffraction
wave vectors with direction and amplitude such that the putative
disproportionated vanadium ASF subtract. Conversely, for the
\emph{forbidden} reflections -as the (1,~0,~0)- the derivative term
remains negligible in front of the combination of the ASF tensors
weighted by its phase factors, and the sum of phase factors of
V-atoms of a given charge state almost cancels out.

By using the multiple scattering theory (MST) we have checked
that, except for the energy shift, the ASF are very little
modified through the transition. Therefore our assumption above is
plainly justified and we use energy shifted HT ASF in the LT
structure. It can also be noticed that the XANES spectra show
little sensitivity to the charge occupancy. The reason is exactly
the same as the non-sensitivity of the (1,~0,~0) reflection: XANES
adds the contributions of the different atoms, the small shifts
due to the CO remain smeared within the experimental
resolution.

We have shown that a charge-ordering transition model
quantitatively explains the resonant scattering experiments in
NaV$_{2}$O$_{5}$. The method presented here has the advantage to
allow for a direct evaluation of the actual ordered charge
disproportionation in the $3d$-transition metal oxides. The charge
ordering concerns a small fraction of an
electron mainly the $d-$orbitals even if a contribution of the
$s-$orbitals cannot be excluded from the present study. This
conclusion, although heavily suspected from the results of other
works, points to reconsider electronic interactions in strongly
correlated electron systems and to the role of small lattice
distortions in accommodating extra charges. Finally with respect
to compounds such as manganites and magnetite, where resonant
diffraction experiments have yielded no sign of CO,
we suspect that the studied diffraction peaks are hardly sensitive
to an actual CO. Our results in NaV$_{2}$O$_{5}$ give
evidence that spectroscopic effects associated to a very small
CO are not at all negligible and that they should be
observable in resonant diffraction experiments if convenient
meaningful peaks are measured and the results contrasted with the
help of appropriate simulations codes.



\begin{thebibliography}{99}

\bibitem{1} J. Zaanen, G. A. Sawatzky and J. W. Allen,
            Phys. Rev. Lett. {\bf 55}, 418 (1985).
\bibitem{2} J. M. D. Coey, M. Viret and S. Von Mol\'ar,
            Adv. Phys. {\bf 48}, 167 (1999).
\bibitem{3} J. P. Wright, J. P. Attfield and P. G. Radaelli,
            Phys. Rev. Lett. {\bf 87}, 266401 (2001).
\bibitem{4} E. J. W. Verwey,
            Nature {\bf 144}, 327 (1939).
\bibitem{5} M. Isobe and Y. Veda,
            J. Phys. Soc. Jpn {\bf 65}, 1178 (1996).
\bibitem{6} D. C. Konigsberger and Roelof Prins,
            {\it X-Ray Absorption: Principles, Applications, Techniques
            of Exafs, Sexafs ans Xanes},Chemical Analysis {\bf 92},
            Wiley Interscience, (1988).
\bibitem{7} J. G. Garcia,  {\it et al.},
             J. Phys.: Condens. Matter {\bf 13}, 3243 (2001).
\bibitem{8} L. Paolasini {\it et al.},
            Phys. Rev. Lett. {\bf 82}, 4719 (1999).
\bibitem{9} J. Goulon {\it et al.},
            Phys. Rev. Lett. {\bf 85}, 4385 (2000).
\bibitem{10} J. G. Garcia,  {\it et al.},
             Phys. Rev. Lett. {\bf 85}, 578 (2000).
\bibitem{11} Y. Murakami {\it et al.},
             Phys. Rev. Lett. {\bf 81}, 582 (1998).
\bibitem{12} M. Benfatto, Y. Joly and C. R. Natoli,
             Phys. Rev. Lett. {\bf 83}, 636 (1999).
\bibitem{13} S. Yu. Ezhov {\it et al.},
             Phys. Rev. Lett. {\bf 83}, 4136 (1999).
\bibitem{14} J. G. Garcia {\it et al.},
             Phys. Rev. B  {\bf 63}, 54110 (2001).
\bibitem{15} S. Grenier {\it et al.},
             Phys. Rev. B  {\bf 65}, 180101 (2002).
\bibitem{16} M. V. Zimmermann {\it et al.},
             Phys. Rev. B {\bf 64}, 195133 (2001).
\bibitem{17} Y. Joly,
             Phys. Rev. B {\bf 63}, 125120 (2001).
\bibitem{18} The program can be freely downloaded at the web adress
             http:/www-cristallo.grenoble.cnrs.fr/simulation.
\bibitem{19} H. Smolinski {\it et al.},
             Phys. Rev. Lett. {\bf 80}, 5164 (1998).
\bibitem{20} H. Sawa {\it et al.},
             J. Phys. Soc. Jpn. {\bf 71}, 385 (2002).
\bibitem{21} P. Blaha,  {\it et al.}
             Comp. Phys. Commun. {\bf 59}, 399 (1990).
\bibitem{22} N. Suaud and M.-B. Lepetit,
             Phys. Rev. Lett. {\bf 88}, 56405 (2002).
\bibitem{joly} Y. Joly,
             J. Synchrotron. Rad. {\bf 10}, 58 (2003).

\end{thebibliography}
\end{document}